\relax

\documentclass{article} 
\usepackage{arxiv} 

\usepackage[hyphens]{url}  
\usepackage{graphicx} 
\usepackage{graphicx}  

\usepackage{dL}
\usepackage{logic}
\usepackage{todonotes}
\usepackage{listings}
\usepackage{xspace}
\usepackage{subfigure}
\usepackage[ruled]{algorithm2e}

\DontPrintSemicolon
\SetKwInOut{Input}{Input}
\SetKwFor{For}{for}{:}{}
\SetKwIF{If}{ElseIf}{Else}{if}{:}{else if}{else:}{}
\SetKwFor{While}{while}{:}{}

\usepackage{prettyref}
\newcommand{\rref}[2][]{\prettyref{#2}}
\newrefformat{model}{Model\,\ref{#1}}
\newrefformat{listing}{Listing\,\ref{#1}}
\newrefformat{line}{line\,\ref{#1}}
\newrefformat{sec}{Section\,\ref{#1}}
\newrefformat{appendix}{Appendix\,\ref{#1}}
\newrefformat{def}{Definition\,\ref{#1}}
\newrefformat{thm}{Theorem\,\ref{#1}}
\newrefformat{ax}{\ref{#1}}
\newrefformat{prop}{Proposition\,\ref{#1}}
\newrefformat{lemma}{Lemma\,\ref{#1}}
\newrefformat{cor}{Corollary\,\ref{#1}}
\newrefformat{ex}{Example\,\ref{#1}}
\newrefformat{tab}{Table\,\ref{#1}}
\newrefformat{fig}{Figure\,\ref{#1}}
\newrefformat{eqn}{Equation~(\ref{#1})}

\newtheorem{example}{Example}

\newcommand{\mname}{\texttt{FOSSIL}\xspace}
\newcommand{\mnameplus}{$\texttt{FOSSIL}^+$\xspace}

\title{Formal Verification of End-to-End Learning in Cyber-Physical Systems: Progress and Challenges}

\author{
Nathan Fulton$^1$, Nathan Hunt$^2$, Nghia Hoang$^1$, Subhro Das$^1$\thanks{NeurIPS Workshop on Safety and Robustness in Decision Making, 2019.} \\
$^1$ MIT-IBM Watson AI Lab, IBM Research\\ 
$^2$ Massachusetts Institute of Technology \\
nathan@ibm.com,  nhunt@mit.edu, \{nghiaht, subhro.das\}@ibm.com
 }

\begin{document}

\maketitle

\begin{abstract}
Autonomous systems -- such as self-driving cars, autonomous drones, and automated trains -- must come with strong safety guarantees. 
Over the past decade, techniques based on formal methods have enjoyed some success in providing strong correctness guarantees for large software systems including operating system kernels, cryptographic protocols, and control software for drones.
These successes suggest it might be possible to ensure the safety of autonomous systems by constructing formal, computer-checked correctness proofs.
This paper identifies three assumptions underlying existing formal verification techniques,
explains how each of these assumptions limits the applicability of verification in autonomous systems,
and summarizes preliminary work toward improving the strength of evidence provided by formal verification.
\end{abstract}

\section{Introduction}

Autonomous systems -- such as cars, planes, and trains -- must come with strong safety guarantees.
These systems are cyber-physical, in the sense that their safety depends crucially upon the way in which their software (``cyber") components interact with their kinetic components.
Cyber-physical systems (CPS) analysis tools can verify the safety of CPS by stating correctness specifications in a formal language and then verifying -- via computer-checked proof -- that safety-critical software components respect these specifications.

Existing approaches toward formally verifying the correctness of cyber-physical systems focus primarily on constructing formal safety proofs about classical low-dimensional models of control systems.
For example, the safety of an adaptive cruise control system might be established by modeling the dynamics of two cars in terms of their positions and velocities and then proving that a control policy preserves safe separation between all cars on the road for any time horizon \cite{LoosPN11}. 
Researchers have employed a similar approach for ensuring the correctness of proposed FAA aircraft collision avoidance protocols \cite{JeanninGKGSZP15}, the European Train Control System \cite{PlatzerQ09}, and quadcopters \cite{RickettsMAGL15}. These proofs are typically constructed and checked using a cyber-physical systems verification tool such as Flow* \cite{ChenAS13}, KeYmaera~X \cite{FultonMQVP15}, or SpaceEx \cite{FrehseGDCRLRGDM11}.

CPS verification tools can provide very strong safety guarantees for cyber-physical systems, but typical techniques for using these tools rely on three assumptions that break down when applying verification techniques to real autonomous systems:
\begin{enumerate}
\item CPS verification techniques assume that a symbolic representation of the state of the world is known a priori. For example, formal CPS models of ground robots typically assume that the system knows the positions of all relevant obstacles, at least within some error bound \cite{MitschP16}.
\item CPS verification techniques assume that a model of all agents' decision making processes is known at design time. The system designer must know what actions the controlled vehicle might take in each state, and must also know what actions \emph{other} agents in the world might take in each state. These models can be nondeterministic, but must be complete in order to obtain meaningful safety results. 
 \item CPS verification techniques assume that the system designer knows a kinematic model that is accurate enough to predict the effect each agents' actions will have on safety-relevant quantities (e.g., position, velocity, etc.).
\end{enumerate}

One common reaction to this list of assumptions is to give up on rationalist approaches toward safety and focus instead on building statistical cases for safety. Unfortunately, a purely statistical approach toward safety certification is intractable. Just as deep reinforcement learning suffers from extreme sample complexity, acquiring sufficient statistical evidence of safety for self-driving cars requires an unrealistic amount of empirical evidence \cite{RANDDriveToSafety}. More importantly, in the purely statistical approach, even slight modifications must trigger a complete recertification. Recent fatalities in both aerospace (737 MAX) and self-driving (Uber Arizona crash) demonstrate the danger of assuming that safety properties of the original system are preserved under even small modifications. If driving our way to safety is the only approach toward obtaining safety guarantees, then fully autonomous systems might never be deployed.

Fortunately, pessimism toward formal methods for autonomous control systems is premature. This paper reviews each of the assumptions listed above and explains how CPS verification techniques can be adapted to address the problem. In the process, we identify a research program based on symbolic reinforcement learning \cite{garnelo2016towards} with domain-specific safety specifications for vision systems and online program synthesis for adjusting modeling assumptions during control.

\section{Background}

We begin by recalling model-based verification and how formal models can be used to provide safety guarantees for reinforcement learning algorithms.

\paragraph{Verification of Model-Based Controllers} In model-based approaches toward safe control, the engineer designs a mathematical model of the system's dynamics. This model is then used to construct a control program that achieves some control objective. The designer ensures the safety of the system by writing down safety specifications and constructing a proof that the control program preserves the desired safety specification.

Model-based verification requires a modeling language. One such language is the language of hybrid programs \cite{Platzer15}, which is a simple programming language for describing the interaction between control software and physical systems. The syntax and informal semantics of HPs are as follows:
\begin{itemize}
\item $\alpha;\beta$ executes $\alpha$ and then executes $\beta$. 
\item $\alpha \cup \beta$ executes either $\alpha$ or $\beta$ nondeterministically.
\item $\alpha^*$ repeats $\alpha$ zero or more times nondeterministically.
\item $x := \theta$ evaluates term $\theta$ and assigns result to $x$. 
\item $x := *$ assigns an arbitrary real value to $x$. 
\item $\{x_1'=\theta_1,...,x_n'=\theta_n \& F\}$ is the continuous evolution of $x_i$ along the solution to the system constrained to a domain defined by $F$.
\item $?F$ aborts if formula $F$ is not true.
\end{itemize}

Safety specifications for hybrid programs are expressed using differential dynamic logic. The formulas of \dL are generated by the grammar:
\[
\varphi,\psi ::= f \sim g ~|~  \varphi \land \psi ~|~ \varphi \lor \psi ~|~ \varphi \limply \psi ~|~ \forall x, \varphi ~|~ \exists x, \varphi ~|~ \dibox{\alpha}\varphi
\]
where $f,g$ are polynomials of real arithmetic, $\sim$ is one of $\{ \le, <, =, >, \ge\}$, and the meaning of $\dibox{\alpha}\varphi$ is that $\varphi$ is true in every state that can be reached by executing the program $\alpha$. \rref{ex:example1} demonstrates how \dL is used to specify the safety property of a pedagogical model of a car approaching a stop sign.

\begin{example}
A pedagogical model of a car at position $x$ approaching a stop sign at position $m$ with velocity $v$ and acceleration $a$. The car's maximum acceleration is $A$ and its maximum braking force is $b$. A time $t$ keeps track of the time between each control step. The evolution domain constraint on the differential equations specifies that the car may not move backward and may choose a new acceleration $a$ at most every $\epsilon$ time steps. The entire formula states that no matter how many control choices the car makes or how much time elapses, the car will not pass the stop sign ($x \le m$)
\begin{lstlisting}[mathescape=true,numbers=left]
$v^2 \le 2b(m-x) \land v \ge 0  \land A \ge 0 \land b>0 \rightarrow$ 
$\big[${
  {a:=-b; $\cup$ ?(2b(m-x) $\ge v^2$+(A+b)(A$\epsilon^2$+2$\epsilon$v)); a:=A;}
  t := 0;
  {x'=v, v'=a, t'=1 & v$\ge$0 & t$\le\epsilon$}
$\}^*\big]$x $\le$ m  
\end{lstlisting} 
\label{ex:example1}
\end{example}

Formulas such as \rref{ex:example1} can be stated and proven in the KeYmaera~X theorem prover \cite{FultonMQVP15} by constructing a proof search tactic \cite{bellerophon}. The particular choice of formal logic and verification technology is not essential to the results presented in this paper. The approach proposed in this paper can be undertaken using temporal logic constraints over B\"{u}chi or Rabin automata, reachability constraints on hybrid automata, or other formalisms.

\paragraph{Formally Constrained Reinforcement Learning} End-to-end reinforcement learning algorithms such as proximal policy optimization \cite{SchulmanWDRK17} learn end-to-end control policies that map directly from sensors to actuation decisions.

Several researchers have suggested approaches toward leveraging logical formulas such as \rref{ex:example1} to provide safety constraints on the action space of reinforcement learning algorithms \cite{AlshiekhBEKNT18,aaai18,HahnPSSTW19,neuralsimplex}. The basic idea is quite simple: a monitor is synthesized from a logical formula and the monitor is used to constrain which actions the RL algorithm may select in each state. For example, an RL agent solving the task considered in \rref{ex:example1} could always choose to brake, but could only accelerate in states where $2b(m-x) \ge v^2+(A+b)(A\epsilon^2+2\epsilon v)$ evaluates to true.

We take this prior work on formally safe RL as a starting point for addressing the three challenges identified in the introduction.

\section{Challenge \#1: Acting Safety Without A Priori Perceptual Knowledge}

CPS verification techniques largely consider the problem of controlling safely given a priori knowledge about the values of safety-relevant quantities. For example, CPS models typically assume error-bounded approximations of positions and velocities for all relevant objects. In the real world, these quantities are only known a posteriori. Properties such as position and velocity are obtained only by analyzing raw sensory inputs (video, LiDAR, radar). 

Consider \rref{ex:example1}. In a real system, the position of the stop sign ($m$) is not part of the system's input. Instead, the system must somehow learn that $m$ is the point at which the car must stop. Autonomous systems that rely on formally constrained RL for safety must correctly map from sensory inputs into the state space in which safety specifications are stated. I.e., the system must correctly couple visual inputs to symbolic states.

\begin{figure}
\centering
  \includegraphics[width=0.6\columnwidth]{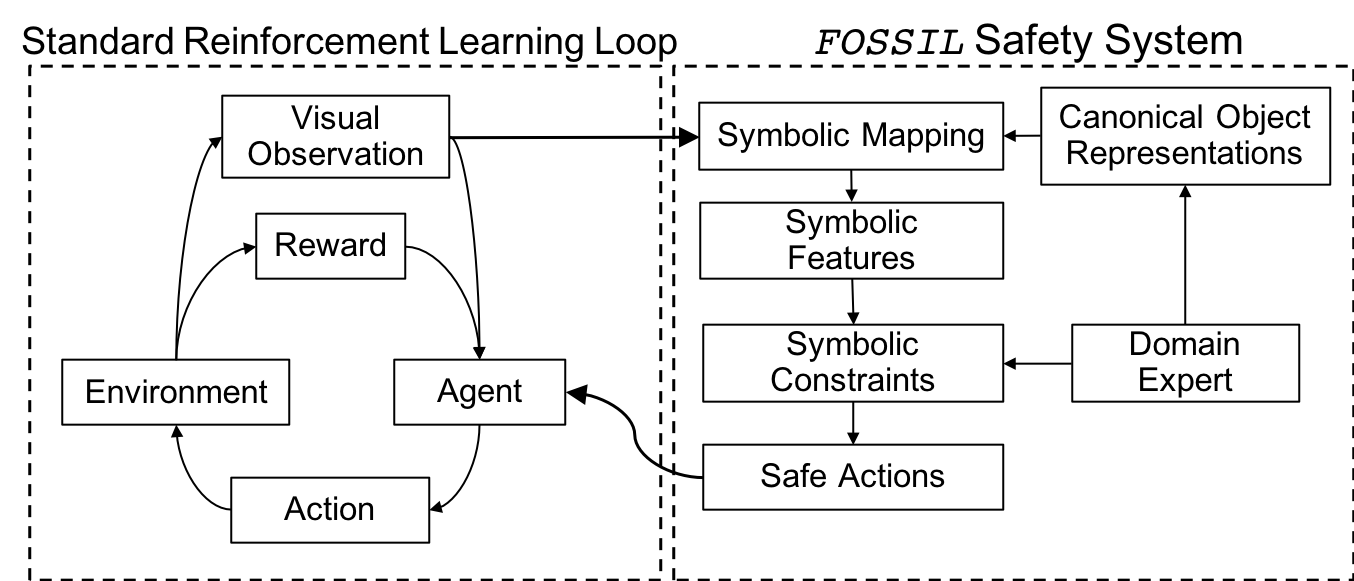}
  \caption{System diagram for the \mname framework.}\label{fig:system_diag}
\end{figure}

We consider this problem in recent work currently under submission. Our approach, called \mname, is a symbolic reinforcement learning algorithm \cite{garnelo2016towards} that uses the symbolic mapping to enforce safety constraints. The algorithm is visualized in \ref{fig:system_diag}. \mname extracts the positions of safety-relevant objects from visual inputs and then enforces safety constraints on the resulting symbolic features. This mapping is achieved via template matching. The symbolic mapping uses quality-aware template mapping \cite{cheng2019qatm,zhou2019objects_centernet,law2018cornernet}.

\begin{figure}
\centering
\includegraphics[width=0.4\columnwidth]{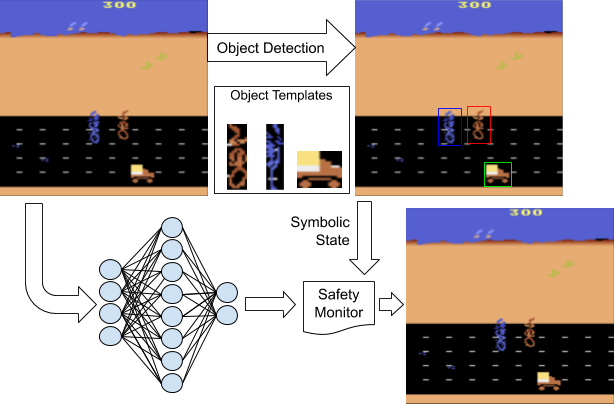}
\caption{\mname instantiated on the Atari Road Runner environment. } \label{fig:rr_instantiation}
\end{figure}

\ref{fig:rr_instantiation} demonstrates how this framework is instantiated on the Atari Road Runner environment. Templates for each safety-relevant object are used to extract the positions of each object, and then a logical safety monitor is used to ensure that the positions of these objects do not violate safety constraints.

Unlike many approaches toward verifiably safe reinforcement learning, \mname learns an end-to-end policy network over visual inputs. Symbolic representations of the system's state are extracted and used to enforce safety constraints, but the system is still learning an end-to-end control policy. This is important because safety constraints describe necessary -- not sufficient -- conditions on adequate control. Autonomous systems must be safe, but many control objectives are not necessary or even possible to capture in a symbolic state space. 

For example, in the Road Runner environment, the road runner must pick up bird seed. 
This bird seed is highly relevant to the reward structure of the game, but is not relevant to the safety objectives of avoiding the car and coyote.
By enforcing constraints in a symbolic state space but still learning end-to-end policies, we are able to optimize for objectives that are not represented in safety models.

\subsection{Future Work on Safe Perception}

\mname explains how to couple visual and symbolic states without requiring a full representation of all safety-relevant features of the symbolic state space. However, \mname does not explain how to ensure that the visual template matching algorithm (or any other vision system) is robust.

Previous work on verification of deep CNNs focuses on adversarial robustness \cite{abs-1811-12395}, which is not the only important specification. Characterizing what it means for a vision system to be correct in general is difficult. However, by taking into account the underlying control problem in which the perception system is being deployed, we can begin stating invariants. For example, Pei et al. verify rotational invariants relevant to steering angle in self-driving systems \cite{abs-1712-01785}. Most important temporal invariants for autonomous systems are temporal: object permanence, physical plausibility, invariance to certain changes in rotation and lighting, etc.

Verifying temporal specifications will require advances in verification tooling. CPS verification tools are excellent for verifying temporal properties, but are designed primarily for analyzing the interaction between the discrete dynamics of control software and the continuous dynamics of physical systems. Their canonical use-case involves low-dimensional but highly descriptive kinematic models of control systems. For example, the dynamics of a car or quadcopter might be modeled using a system of 10-15 differential equations that directly reference the positions and velocities of the controlled vehicle and any relevant obstacles. Analysis of high-dimensional systems is also sometimes possible, but only for affine systems and for certain types of specifications \cite{bak2019hscc}. Verification approaches for CNNs do scale to large dimensions, but do not consider the temporal nature of classification tasks in autonomous systems.

\section{Challenges \#2 and \#3: Predicting Behaviors and their Effects}

Modeling the world is difficult. Even simple control problems often require substantial effort to accurately model. Most approaches toward safe reinforcement learning enforce constraints on the action space, such as the assertion on line 3 of \rref{ex:example1}. These safety constraints contain, often implicitly, assumptions about how the world behaves. For example, line 3 of \rref{ex:example1} is directly related to the solution to the system of differential equations on line 5 of \rref{ex:example1}. If those differential equations do not adequately capture the behavior of the car under control, then the safety constraint given on line 3 is wrong.

Autonomous systems that rely on state-action constraints to ensure safety should reify the modeling assumptions underlying these constraints. These modeling assumptions should be continually checked at runtime to ensure that the safety system is not enforcing constraints based upon erroneous assumptions. This principle raises an important question: what should the system do when detecting a mismatch between expectation and reality?

One answer to this question, explored in \cite{aaai18}, is that the system should attempt to ``patch" model mismatches by using the distance between expectation and reality as an objective function. Although this approach seems to work well in cases where error is small in magnitude and intermittent, it is not a particularly principled approach. 

\begin{figure}
\centering
\includegraphics[width=0.9\columnwidth]{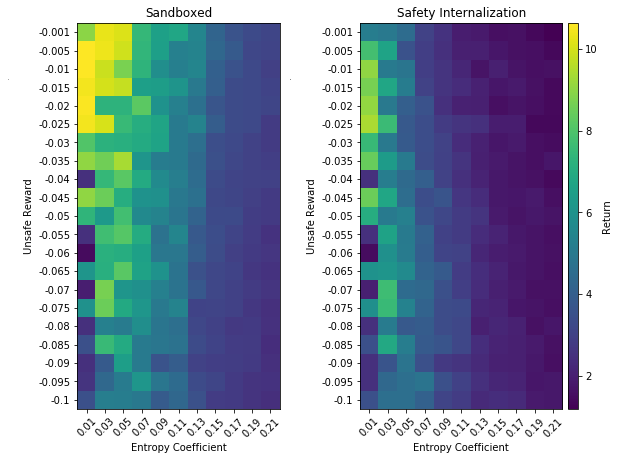}
\caption{Safety generalization ability when penalizing unsafe action attempts. Penalization (``Unsafe Reward" $\not =$ 0) degrades system performance.}
  \label{fig:unsafepen}
\end{figure}

Several researchers have suggested that reinforcement learning algorithms should \emph{internalize} knowledge about unsafe actions by penalizing hypothetically unsafe behavior whenever the control policy suggests an action that is rejected by a formal safety guard. We explore this proposal in the context of \mname and find that doing so decreases overall performance and does not succeed in guiding the system away from unsafe states. These results are illustrated in \rref{fig:unsafepen}.

Thus, we suggest a more principled approach: the system should attempt to understand the cause of its incorrect predictions, construct a new model of the world that corrects for these modeling inaccuracies, and then synthesize new provably correct control logic. We consider this approach in the context of \mname. We introduce a modified algorithm, \mnameplus, that is told about its action space and is them left to learn the effects its actions have on the world. We find that \mnameplus constructs far better action space constraints than those provided based upon a priori modeling assumptions. Although these experiments are exploratory, they suggest that verification-preserving program synthesis is the most promising approach toward addressing modeling inaccuracies in cyber-physical systems.

\begin{figure}
\centering
\includegraphics[width=0.5\columnwidth]{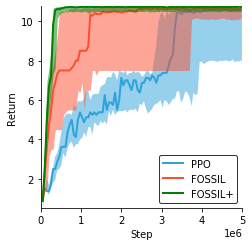}  
\caption{Cumulative reward during training for \mname and \mnameplus.}
\label{fig:cumre}
\end{figure}

Our initial successes with \mnameplus and initial failures with optimization-based adaptation techniques suggests that online program synthesis should play a more prominent role in design of safe autonomous systems.

\section{Conclusion}

Current verification techniques are not sufficient for designing safe autonomous cyber-physical systems. Existing verification techniques assume an accurate model of the world is known a priori and do not consider the problem of adapting to observed modeling inaccuracies. Our recent work on \mname demonstrates how to integrate symbolic safety constraints into an end-to-end reinforcement learning system without relying on an oracle for obtaining a symbolic representation of visual inputs. In conclusion, we suggest two broad areas for future work on verifiably safe learning for cyber-physical systems: verification technology for checking temporal specifications of computer vision systems, and  efficient verification-preserving program synthesis algorithms for hybrid programs.

\fontsize{9.5pt}{10.5pt} \selectfont
\bibliographystyle{plain}
\bibliography{manuscript_SafeSymbolicRL}

\begin{thebibliography}{10}

\bibitem{AlshiekhBEKNT18}
Mohammed Alshiekh, Roderick Bloem, R{\"{u}}diger Ehlers, Bettina
  K{\"{o}}nighofer, Scott Niekum, and Ufuk Topcu.
\newblock Safe reinforcement learning via shielding.
\newblock In {\em Proceedings of the Thirty-Second {AAAI} Conference on
  Artificial Intelligence ({AAAI} 2018)}, 2018.

\bibitem{bak2019hscc}
Stanley Bak, Hoang-Dung Tran, and Taylor~T. Johnson.
\newblock Numerical verification of affine systems with up to a billion
  dimensions.
\newblock In {\em Proceedings of the 22Nd ACM International Conference on
  Hybrid Systems: Computation and Control}, HSCC '19, pages 23--32, New York,
  NY, USA, 2019. ACM.

\bibitem{abs-1811-12395}
Akhilan Boopathy, Tsui{-}Wei Weng, Pin{-}Yu Chen, Sijia Liu, and Luca Daniel.
\newblock Cnn-cert: An efficient framework for certifying robustness of
  convolutional neural networks.
\newblock {\em CoRR}, abs/1811.12395, 2018.

\bibitem{ChenAS13}
Xin Chen, Erika {\'{A}}brah{\'{a}}m, and Sriram Sankaranarayanan.
\newblock Flow*: An analyzer for non-linear hybrid systems.
\newblock In {\em Computer Aided Verification - 25th International Conference
  ({CAV} 2013)}, pages 258--263, 2013.

\bibitem{cheng2019qatm}
Jiaxin Cheng, Yue Wu, Wael Abd-Almageed, and Premkumar Natarajan.
\newblock Qatm: Quality-aware template matching for deep learning.
\newblock In {\em Proceedings of the IEEE Conference on Computer Vision and
  Pattern Recognition}, pages 11553--11562, 2019.

\bibitem{FrehseGDCRLRGDM11}
Goran Frehse, Colas~Le Guernic, Alexandre Donz{\'{e}}, Scott Cotton, Rajarshi
  Ray, Olivier Lebeltel, Rodolfo Ripado, Antoine Girard, Thao Dang, and Oded
  Maler.
\newblock {SpaceEx}: Scalable verification of hybrid systems.
\newblock In {\em Computer Aided Verification - 23rd International Conference
  ({CAV} 2011)}, pages 379--395, 2011.

\bibitem{bellerophon}
Nathan Fulton, Stefan Mitsch, Brandon Bohrer, and Andr{\'{e}} Platzer.
\newblock Bellerophon: Tactical theorem proving for hybrid systems.
\newblock In Mauricio Ayala{-}Rinc{\'{o}}n and C{\'{e}}sar~A. Mu{\~{n}}oz,
  editors, {\em Interactive Theorem Proving - 8th International Conference
  ({ITP} 2017)}, volume 10499 of {\em {LNCS}}, pages 207--224. Springer, 2017.

\bibitem{FultonMQVP15}
Nathan Fulton, Stefan Mitsch, Jan-David Quesel, Marcus V{\"o}lp, and Andr{\'e}
  Platzer.
\newblock {KeYmaera X}: An axiomatic tactical theorem prover for hybrid
  systems.
\newblock In {\em CADE}, volume 9195, pages 527--538, 2015.

\bibitem{aaai18}
Nathan Fulton and Andr{\'e} Platzer.
\newblock Safe reinforcement learning via formal methods: Toward safe control
  through proof and learning.
\newblock In Sheila McIlraith and Kilian Weinberger, editors, {\em Proceedings
  of the Thirty-Second {AAAI} Conference on Artificial Intelligence ({AAAI}
  2018)}, pages 6485--6492. {AAAI} Press, 2018.

\bibitem{garnelo2016towards}
Marta Garnelo, Kai Arulkumaran, and Murray Shanahan.
\newblock Towards deep symbolic reinforcement learning.
\newblock {\em arXiv preprint arXiv:1609.05518}, 2016.

\bibitem{HahnPSSTW19}
Ernst~Moritz Hahn, Mateo Perez, Sven Schewe, Fabio Somenzi, Ashutosh Trivedi,
  and Dominik Wojtczak.
\newblock Omega-regular objectives in model-free reinforcement learning.
\newblock In Tom{\'{a}}s Vojnar and Lijun Zhang, editors, {\em Tools and
  Algorithms for the Construction and Analysis of Systems ({TACAS} 2019)},
  volume 11427 of {\em Lecture Notes in Computer Science}, pages 395--412.
  Springer, 2019.

\bibitem{JeanninGKGSZP15}
Jean{-}Baptiste Jeannin, Khalil Ghorbal, Yanni Kouskoulas, Ryan Gardner, Aurora
  Schmidt, Erik Zawadzki, and Andr{\'e} Platzer.
\newblock A formally verified hybrid system for the next-generation airborne
  collision avoidance system.
\newblock In Christel Baier and Cesare Tinelli, editors, {\em TACAS}, volume
  9035 of {\em LNCS}, pages 21--36. Springer, 2015.

\bibitem{RANDDriveToSafety}
Nidhi Kalra and Susan~M. Paddock.
\newblock {\em Driving to Safety: How Many Miles of Driving Would It Take to
  Demonstrate Autonomous Vehicle Reliability?}
\newblock RAND Corporation, 2016.

\bibitem{law2018cornernet}
Hei Law and Jia Deng.
\newblock Cornernet: Detecting objects as paired keypoints.
\newblock In {\em Proceedings of the European Conference on Computer Vision
  (ECCV)}, pages 734--750, 2018.

\bibitem{LoosPN11}
Sarah~M. Loos, Andr{\'{e}} Platzer, and Ligia Nistor.
\newblock Adaptive cruise control: Hybrid, distributed, and now formally
  verified.
\newblock In {\em {FM} 2011: Formal Methods - 17th International Symposium on
  Formal Methods, Limerick, Ireland, June 20-24, 2011. Proceedings}, pages
  42--56, 2011.

\bibitem{MitschP16}
Stefan Mitsch and Andr{\'e} Platzer.
\newblock {ModelPlex}: Verified runtime validation of verified cyber-physical
  system models.
\newblock {\em Form. Methods Syst. Des.}, 49(1):33--74, 2016.

\bibitem{abs-1712-01785}
Kexin Pei, Yinzhi Cao, Junfeng Yang, and Suman Jana.
\newblock Towards practical verification of machine learning: The case of
  computer vision systems.
\newblock {\em CoRR}, abs/1712.01785, 2017.

\bibitem{neuralsimplex}
Dung Phan, Nicola Paoletti, Radu Grosu, Nils Jansen, Scott~A. Smolka, and
  Scott~D. Stoller.
\newblock Neural simplex architecture.
\newblock {\em CoRR}, abs/1908.00528, 2019.

\bibitem{Platzer15}
Andr{\'e} Platzer.
\newblock A uniform substitution calculus for differential dynamic logic.
\newblock In {\em CADE}, 2015.

\bibitem{PlatzerQ09}
Andr{\'e} Platzer and Jan-David Quesel.
\newblock {European Train Control System}: A case study in formal verification.
\newblock In Karin Breitman and Ana Cavalcanti, editors, {\em ICFEM}, volume
  5885 of {\em LNCS}, pages 246--265. Springer, 2009.

\bibitem{RickettsMAGL15}
Daniel Ricketts, Gregory Malecha, Mario~M. Alvarez, Vignesh Gowda, and Sorin
  Lerner.
\newblock Towards verification of hybrid systems in a foundational proof
  assistant.
\newblock In {\em 13. {ACM/IEEE} International Conference on Formal Methods and
  Models for Codesign, {MEMOCODE} 2015, Austin, TX, USA, September 21-23,
  2015}, pages 248--257. {IEEE}, 2015.

\bibitem{SchulmanWDRK17}
John Schulman, Filip Wolski, Prafulla Dhariwal, Alec Radford, and Oleg Klimov.
\newblock Proximal policy optimization algorithms.
\newblock {\em CoRR}, abs/1707.06347, 2017.

\bibitem{zhou2019objects_centernet}
Xingyi Zhou, Dequan Wang, and Philipp Krahenbuhl.
\newblock Objects as points.
\newblock {\em arXiv preprint arXiv:1904.07850}, 2019.

\end{thebibliography}

\end{document}